\documentclass[aps,pra,twocolumn,superscriptaddress,showpacs]{revtex4-2}
\usepackage{indentfirst}
\usepackage{graphicx}
\usepackage{dcolumn}
\usepackage{bm}
\usepackage{amsmath}
\usepackage{amssymb}
\usepackage{latexsym}
\usepackage{epsfig}
\usepackage{amsbsy}
\usepackage{array}
\usepackage{amssymb}
\usepackage{setspace}
\usepackage{bm}
\usepackage{endnotes}
\usepackage{indentfirst}
\begin{document}

\preprint{Physical Review A}

\title{What is the meaning of physical quantity $\nu$ in the expression of photon energy $h\nu$?}

\author{Xingchu Zhang}
\affiliation{School of Physics and Information Engineering, Guangdong University of Education, Guangzhou 510303, China }

\author{Weilong She }
\email[]{shewl @mail.sysu.edu.cn}
\thanks{co-first author.}
\affiliation{School of Physics, Sun Yat-Sen University, Guangzhou 510275, China.}


\begin{abstract}
It is well known that, for an incident light of not so high intensity and in a certain range of frequency, the stopping voltage of photoelectric effect is independent of the intensity but dependent on the frequency of the light, which is described by the equation  $V = h\nu /e - {W_0}/e$, where $V$ is the stopping voltage, $h$ is the Planck constant, $\nu$ is the frequency of incident light, $e$ is the electron charge, and  $W_0$ is the work function. It means that the larger the frequency of incident light, the higher the stopping voltage is. However, the present experiment finds that for a non-monochromatic incident light, the stopping voltage is not determined by the maximum frequency component of the incident light, but by the maximum center frequency of all the wave train components (with different center frequencies) involved in the incident light, that is to say, in the photon energy expression $h\nu$, physical quantity $\nu$ does not refer to the frequency of a monochromatic light, but represents the center frequency of a wave train spectrum. The spectral bandwidth of a wave train component can be as large as 122 nm in visible and near-infrared region. This should arouse more attention in the study of energy exchange between light and matter.
\end{abstract}


\maketitle

\section{Introduction}
The concept of light quantum (photon) is one of the basic concepts in modern physics. It was proposed by Einstein\cite{Einstein1}, and was gradually accepted after being confirmed by photoelectric effect\cite{Millikan2} and Compton scattering experiments\cite{Compton3}. Now the concept is widely used in particle physics\cite{Eidelman4}, the photon rest mass detection\cite{Luo5}, photon bunching or antibunching\cite{Wolf6,Press7}, photon-by-photon scattering\cite{Burke8,ATLAS9}, photon coherent absorption\cite{Roger10}, two-photon spontaneous emission\cite{Muniz11}, photon entangled pairs generation\cite{Kwiat12,Barreiro13}, quantum entanglement\cite{Ecker14,Sun15} and other optical processes. According to Einstein’s theory, a light quantum is with the energy of 
\begin{eqnarray}
E = h\nu. 
\label{eq:energy}
\end{eqnarray}
This is a well-known formula in optoelectronics\cite{Wilson16} and spectroscopy\cite{Wolfgang17} and is usually used for the fundamental process of energy exchange between light and matter. Applying the concept of light quantum to the photoelectric effect, Einstein gave his famous equation ${E_k} = h\nu  - {W_0}$\cite{Einstein1}, where $ E_k$ is the maximum kinetic energy of the photoelectrons. Now let’s focus our attention on the stopping voltage $V$, which can be derived from the Einstein equation and reads
\begin{eqnarray}
V = \frac{{h\nu }}{e} - \frac{{{W_0}}}{e}. 
\label{eq:energy}
\end{eqnarray}
However, the photon energy $h\nu$ in the expression still needs more understanding and exploration. The question is, does physical quantity $\nu$ here represent the frequency of a monochromatic light wave, or does it represent the center frequency of a wave train spectrum? If regard physical quantity $\nu$  as the frequency of a monochromatic light wave, according to Eq.(2), for a non-monochromatic incident light with showy bandwidth, the measured value of the stopping voltage should correspond to the side frequency of maximum value, because the stopping voltage is determined by the photoelectron with the maximum kinetic energy, and in this case, the maximum kinetic energy should be related to the maximum frequency of light wave. But is it the case? If physical quantity $\nu$ is the center frequency of a wave train spectrum, then how to determine it when the incident light is a non-monochromatic one? And what is the spectral bandwidth corresponding to the center frequency $\nu$ ?  Is there an upper bound for the bandwidth? As far as we know, there is no existing answer to these questions. It can only be judged by experiment. Our experiment shows that, for a non-monochromatic incident light, the stopping voltage of photoelectric effect is not determined by the maximum frequency component of the incident light, but by the maximum center frequency of the wave train components (with different center frequencies) involved in the incident light. That is to say, in the expression $h\nu$ of Eq.(2), physical quantity $\nu$ does not represent the frequency of a monochromatic light, but represents the center frequency of a wave train spectrum. The experiment also finds that in the visible and near infrared regions, the spectral bandwidth of a wave train component can be as large as 122 nm. The experiment is carried out by four steps:\\
(1) Use discrete quasi-monochromatic lights of wavelengths 365, 404.7, 435.8, 546.1, 577, 589.3 and 632.8 nm for experiment. It is found that the stopping voltage obeys Eq.(2) if these lights are used respectively; and the stopping voltage is determined by the light of maximum frequency when multiple quasi-monochromatic lights are incident onto the photoelectric tube simultaneously.\\
(2) Use non-monochromatic lights ($\lambda_0 = 560$ nm, $\Delta \lambda  = 10$ nm), ($\lambda_0 = 404.7$ nm, $\Delta \lambda  = 39$ nm), ($\lambda_0 = 550$ nm, $\Delta \lambda  = 40$ nm), ($\lambda_0 = 540$ nm, $\Delta \lambda  = 98$ nm) and ($\lambda_0 = 653$ nm, $\Delta \lambda  = 108$ nm), respectively for experiment, where $\lambda_0$ is the center wavelength and $\Delta \lambda$ the bandwidth. It is found that the stopping voltage is determined by the center wavelength (frequency) of each non-monochromatic incident light.\\ 
(3) Use non-monochromatic lights ($\lambda_0 = 600$ nm, $\Delta \lambda  = 281$ nm), ($\lambda_0 = 659$ nm, $\Delta \lambda  = 138$ nm), ($\lambda_0 = 535$ nm, $\Delta \lambda  = 107.2$ nm) and ($\lambda_0 = 413$ nm, $\Delta \lambda  = 62.6$ nm) respectively for experiment. It is found that the stopping voltage is no longer determined by the center wavelengths of each non-monochromatic incident light, but corresponds to a wavelength shorter than the center one.\\
(4) Combine non-monochromatic light ($\lambda_0 = 653$ nm, $\Delta \lambda  = 108$ nm) with quasi-monochromatic light 589.3 nm from a sodium lamp, or 577 nm, or 546 nm from a mercury lamp for experiment. It is found that the stopping voltage is determined by each added quasi-monochromatic light.  The similar phenomenon is found by combining non-monochromatic light ($\lambda_0 = 600$ nm, $\Delta \lambda  = 281$ nm) with quasi-monochromatic light 365 nm, or 404.7 nm, or 435.8 nm from a mercury lamp for experiment.

\section{Experiment and Results}

In the first step of experiment, we use He-Ne laser, sodium lamp and mercury lamp as light sources, and use a ZKY-GD-4 experimental setup of photoelectric effect (the wavelength response is from 340 nm to 700 nm) as a detector to detect the stopping voltage. The spectral lines used are 365, 404.7, 435.8, 546.1, 577, 589.3 and 632.8 nm, respectively. The experimental result is shown in Fig.1, 
\begin{figure}
\centering
\includegraphics[width=3.5in]{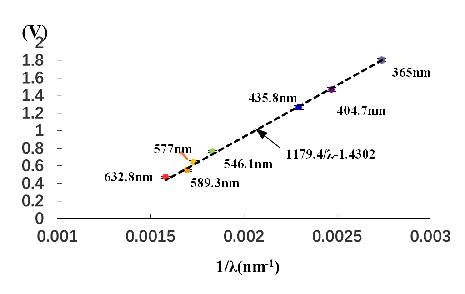}
\caption{\label{fig1}Use discrete quasi-monochromatic lights of wavelengths 365, 404.7, 435.8, 546.1, 577, 589.3 and 632.8 nm respectively for experiment. It is found that the stopping voltage obeys Eq.(3).}
\end{figure}
where the points with wavelength are the experimental data, and the dashed straight line is the result obtained by fitting the experimental data with the deformation function of Eq.(2),i.e.,
\begin{eqnarray}
V = Kx + b, 
\label{eq:energy}
\end{eqnarray}

where $x=1/\lambda$, $\lambda$ is the wavelength in nm. The fitting parameters are $K=1179.4$ Vnm and $b=-1.4302 $ V, respectively. Not surprisingly, the experimental points are almost on the straight line. It confirms that the experimental device works normally and reliably. We then let the light involving all the spectral lines($\ge 365$ nm) of the mercury lamp be incident onto the photoelectric tube simultaneously. The stopping voltage is found to be 1.7805 V, which is close to 1.7912 V measured by using single spectral line 365 nm, but slightly lower. The value of slightly lower than 1.7912 V is due to the shielding effect of the low-speed electrons (that cannot reach the negative electrode of the photoelectric tube) produced by the long-wavelength lights. In order to reduce the effect, the entrance aperture of the photoelectric tube should be chosen as small as possible. In the experiment, we choose an aperture of 2 mm diameter. This experiment seems to prove that the physical quantity $\nu$ in Eq.(2) represents the frequency of a monochromatic light, since the stopping voltage is really determined by the maximum frequency component of the incident light although the incident light involves many of quasi-monochromatic components.  

Now there is such a question: If the incident light is a non-monochromatic one, furthermore, with a wider continuous spectrum, is the stopping voltage determined by the shortest wavelength in the spectrum? To answer this question, we started the second step of experiment. We use a bromine tungsten lamp as the light source, and use five filters to get several of non-monochromatic lights of different bandwidths for experiment. The experimental results are shown in Fig.2. 
\begin{figure*}
\centering
\includegraphics[width=7in]{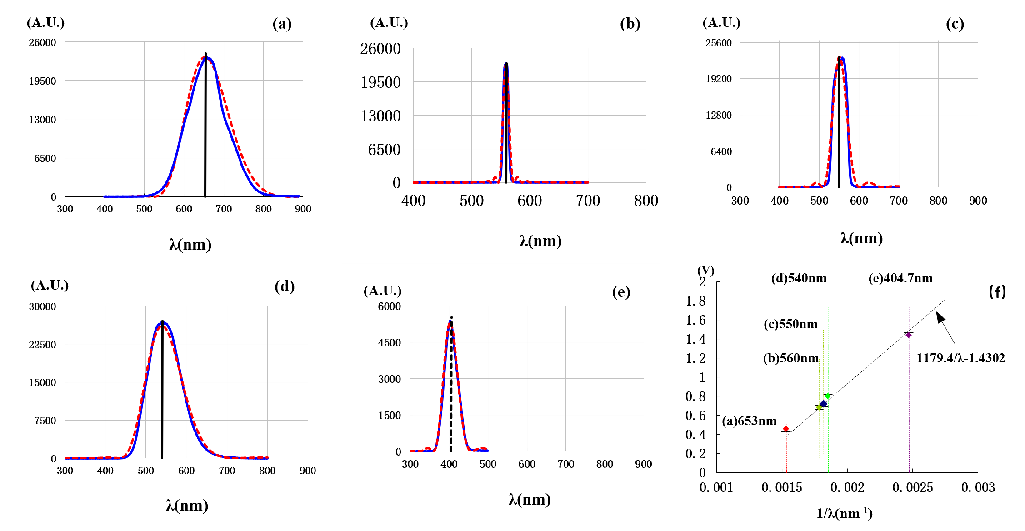}
\caption{\label{fig2}Use non-monochromatic lights ($\lambda_0 = 653$ nm, $\Delta \lambda  = 108$ nm)(a), ($\lambda_0 = 560$ nm, $\Delta \lambda  = 10$ nm)(b), ($\lambda_0 = 550$ nm, $\Delta \lambda  = 40$ nm)(c), ($\lambda_0 = 540$ nm, $\Delta \lambda  = 98$ nm)(d)  and ($\lambda_0 = 404.7$ nm, $\Delta \lambda  = 39$ nm)(e) respectively for experiment, where $\lambda_0$ is the center wavelength and $\Delta \lambda$ is the bandwidth. It is found that the stopping voltage is determined by the center wavelength (frequency) of each non-monochromatic incident light[see (f)].}
\end{figure*}
The solid blue lines of (a) - (e) are the spectra of various non-monochromatic lights, in which, the center wavelengths and bandwidths are respectively: (a)  $\lambda_0 = 653$ nm, $\Delta \lambda  = 108$ nm, (b) $\lambda_0 = 560$ nm, $\Delta \lambda  = 10$ nm, (c) $\lambda_0 = 550$ nm, $\Delta \lambda  = 40$ nm, (d) $\lambda_0 = 540$ nm, $\Delta \lambda  = 98$ nm and (e) $\lambda_0 = 404.7$ nm, $\Delta \lambda  = 39$ nm. The red dashed lines in the figure are the fitting ones (the fitting function will be given below). The experiment finds that the stopping voltage is determined by the center wavelength rather than the shortest wavelength of each spectrum, although the spectral bandwidths of incident lights vary greatly, ranging from 10 nm to 108 nm. Figure 2(f) shows the stopping voltages observed. The black dashed straight line in Fig.2(f) is from the function $V = 1179.4x - 1.4302$. And the color vertical dashed lines marked with the wavelength correspond to the black vertical dashed lines in Fig.2 (a) - (e). The above two steps of experiments show that physical quantity $\nu$ in Eq.(2) represents the center frequency of the spectrum of incident light than the frequency of the monochromatic component involved in the incident light. The spectral bandwidth of incident light can be very narrow, like that of He-Ne laser and those of mercury lamp lines; and also can be very wide, like those shown in Fig.2, especially in Fig.2(a), being 108 nm.

Now there is still another question: Is the stopping voltage always determined by the center wavelength of the spectrum of incident light? To answer this question, we carried out the third step of experiment. At first, we directly use the radiation of bromine tungsten lamp for experiment; then use other filters, with wider transmission spectra than those of Figs2.(a) and (e), to get the lights of wider spectra from the bromine tungsten lamp for experiment. The results are shown in Fig.3. Figure 3(a) is the spectrum of the bromine tungsten lamp with a center wavelength at 600 nm and a spectral bandwidth of $\Delta \lambda  = 281$ nm. And Figs.3(b) and (c) are those got by using the filters, which are respectively: (b) $\lambda_0 = 659$ nm and $\Delta \lambda  = 138$ nm,  (c) $\lambda_0 = 413$ nm and $\Delta \lambda  = 62.6$ nm. The solid blue lines of (a) - (c) are the measured spectra of incident lights, and the dashed red lines are the fitted ones (the fitting function will be given below). Note that in (b) and (c), the blue and red lines are almost overlapped. In Figs.3(a) - (c), the colored vertical lines indicate the wavelengths corresponding to the stopping voltages, which are 439 nm [corresponding to 1.256 V](a), 562 nm [corresponding to 0.668 V] (b) and 400 nm [corresponding to 1.518 V] (c), respectively. They are no longer the center wavelengths of these spectra, but are at those positions shorter than the center wavelengths. This is completely different from the situation shown in Figs.2. It should be noticed that the spectral bandwidths of Fig.2(a) and Fig.3(b) are different, being 108 nm and 138nm respectively; and this difference has led to the significant difference of stopping voltage: in Fig.2(a), the stopping voltage corresponds to the center wavelength of incident light, while in Fig.3(b), the stopping voltage is determined by a wavelength shorter than the center one. 
\begin{figure*}
\centering
\includegraphics[width=7in]{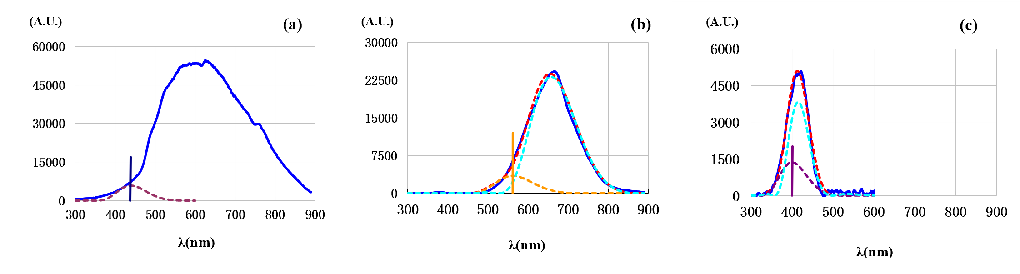}
\caption{\label{fig3}Use non-monochromatic lights ($\lambda_0 = 600$ nm, $\Delta \lambda  = 281$ nm)(a), ($\lambda_0 = 659$ nm, $\Delta \lambda  = 138$ nm)(b), and ($\lambda_0 = 413$ nm, $\Delta \lambda  = 62.6$ nm)(c) respectively for experiment. It is found that the stopping voltage is no longer determined by the center wavelengths of each non-monochromatic incident light, but corresponds to a wavelength shorter than the center one(see the colored vertical lines)}.
\end{figure*} 

In the fourth step of experiment, we combine non-monochromatic light of Fig.2(a) ($\lambda_0 = 653$ nm, $\Delta \lambda  = 108$ nm) with quasi-monochromatic light 589.3 nm from a sodium lamp, or 577 nm, or 546 nm from a mercury lamp for experiment. The combination of lights is done by using a splitting prism of no polarization selectivity. Each quasi-monochromatic light goes through the prism and the non-monochromatic light is reflected by the prism then they combine together and go to the photoelectric tube. The results are shown in Fig.4. Figure 4(a) is the spectrum of Fig.2(a) combined with 589.3 nm, 577 nm and 546 nm lines. The colored vertical lines indicate the positions of quasi-monochromatic light (the red one indicates center wavelength of Fig.2 (a)). Figure 4(c) shows the relationship between stopping voltage and wavelength. One sees that each stopping voltage is determined by the quasi-monochromatic light and is slightly lower than that when the quasi-monochromatic light is used alone (this is also due to the shielding effect of the low-speed electrons produced by the long-wavelength light). It should be mentioned that when the added quasi-monochromatic light is removed, the stopping voltage comes back to that corresponding to the center wavelength of Fig.2(a). Note that in the second step of experiment, although the light in Fig.2(a) involves three components of 589.3 nm, 577 nm and 546 nm, the stopping voltage hardly reflects the effect of these components there. We further carry out another combination experiment. We combine non-monochromatic light of Fig.3(a) ($\lambda_0 = 600$ nm, $\Delta \lambda  = 281$ nm) with quasi-monochromatic light 365 nm, or 404.7 nm, or 435.8 nm from a mercury lamp for experiment. Figure 4(b) is the spectrum of Fig.3(a) combined with 365 nm, 404.7 nm and 435.8 nm lines. And the stopping voltages are also shown in Fig.4(c), where RBTL represents the radiation of the bromine tungsten lamp. Obviously each stopping voltage is determined by the added quasi-monochromatic light. Although the light used in Fig.3(a) involves three components of 365 nm, 404.7 nm and 435.8 nm, the stopping voltage also hardly reflects the effect of these components there.  
\begin{figure*}
\centering
\includegraphics[width=7in]{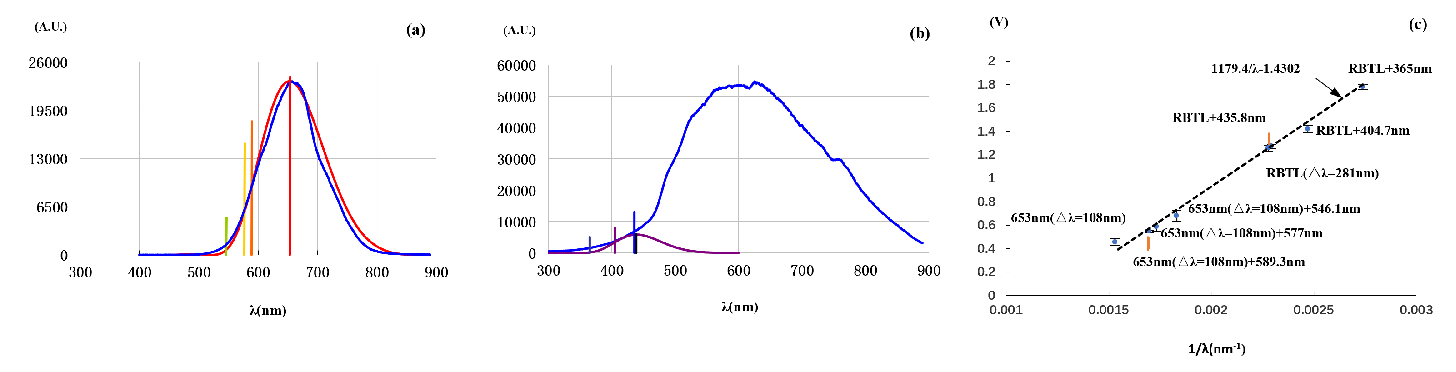}
\caption{\label{fig4}Combine non-monochromatic light ($\lambda_0 = 653$ nm, $\Delta \lambda  = 108$ nm) with quasi-monochromatic light 589.3 nm from a sodium lamp, or 577 nm, or 546 nm from a mercury lamp for experiment. It is found that the stopping voltage is determined by each added quasi-monochromatic light [see (c)]. The similar phenomenon [see (c)] is found by combining non-monochromatic light ($\lambda_0 = 600$ nm, $\Delta \lambda  = 281$ nm) with quasi-monochromatic light 365 nm, or 404.7 nm or 435.8 nm from a mercury lamp for experiment. Where (a) and (b) show the spectra of two non-monochromatic lights and the added quasi-monochromatic lights; RBTL represents the radiation of the bromine tungsten lamp.}
\end{figure*} 

\section{Discussion}
Let's discuss the above experimental results. It is well known that the stopping voltage of the photoelectric effect is independent of the light intensity, and corresponds to the maximum kinetic energy of photoelectrons. If the physical quantity $\nu$ in Eq.(2) refers to the frequency of a monochromatic light, then the measured value of the stopping voltage should always correspond to the smallest wavelength (or the maximum frequency) of the spectrum of incident light, but the experimental results are not the case. In fact, the stopping voltage may be determined by the center wavelength of the spectrum of incident light, as the cases of Fig.2, or determined by a wavelength component shorter than the center one of the incident light, as the case of Fig.3. Let's look at Fig.2(a) and Fig.4(a) again. The spectrum of Fig.2(a) clearly involves the wavelengths of 589.3 nm, 577 nm, and 546 nm, but they do not play a decisive role in generating stopping voltage; while on the basis of Fig.2(a), when adding any of extra quasi-monochromatic lights with center wavelengths 589.3 nm, 577 nm and 546 nm, it immediately plays a decisive role in generating stopping voltage (shown in Figs.4 (a) and (c)). Similar phenomenon can be found in Fig.3(a), Figs.4 (b) and (c). How to understand present experiment?  Let's try to explain it with discrete wavelet structure theory of classical plane light waves\cite{Zhang18}, which is consistent with the coherent state theory of quantized light field \cite{Glauber19}(see \cite{20}). According to the theory, a basic plane light wave with limited coherence length can be described by following wave train with discrete wavelet structure 
\begin{widetext}
\begin{eqnarray}
{E_k}(z - ct) = \frac{{{E_{k0}}}}{2}{e^{ik(z - ct)}}\sum\limits_{r =  - n}^{n + 1} {{e^{\frac{{ - {{[(z - ct) - r{\lambda _0} + {\lambda _0}/2]}^2}}}{s}}}} ,(n = 1,2,3, \cdots ).
\label{eq:energy}
\end{eqnarray}
\end{widetext}
Here $c$ is the speed of light in vacuum, $\lambda _0$ is the center wavelength, $E_{k0}$ is a constant, and  ${\omega _0} = 2\pi c/{\lambda _0} = 2\pi {\nu _0}$ is the center angular frequency, $k = {\omega _0}/c$, $s = {({\lambda _0}/0.8862319204)^2}$(which is a parameter for decomposing the wave train as a series of discrete wavelets)\cite{Zhang18}. Note that with a large $n$, Eq.(4) can also describe a quasi-monochromatic plane light wave, for example any of spectral lines used in the first step of experiment. The energy of this wave train is discrete\cite{Zhang18}. And along $\vec k$ and through a cross-section ${\lambda _0^2} $, the energy that can be devolved to a medium is also discrete, being $j{p_{0k}}{\omega _0}(j = 1,2,3, \cdots ,n) $, where $p_{0k}$ is a constant. The spectral distribution of the wave train is
\begin{widetext}
\begin{eqnarray}
F(\omega ) = {\left| {\frac{{{E_{k0}}}}{2}{e^{ - \frac{{s{{(\omega  - {\omega _0})}^2}}}{{4{c^2}}}}}\sum\limits_{r =  - n}^{n + 1} {{e^{ - i(\omega  - {\omega _0})[r{\lambda _0}/c - {\lambda _0}/2c]}}} } \right|^2}, (n = 1,2,3, \cdots ).
\label{eq:energy}
\end{eqnarray}
If $E_{k0}/2$ is taken as one unit and the spectrum is represented in wavelength, it becomes
\begin{eqnarray}
F(\lambda ) = {\left| {{e^{ - {\pi ^2}s{{(1/\lambda  - 1/{\lambda _0})}^2}}}\sum\limits_{r =  - n}^{n + 1} {{e^{ - i2\pi (1/\lambda  - 1/{\lambda _0})[r{\lambda _0} - {\lambda _0}/2]}}} } \right|^2}, (n = 1,2,3, \cdots ).
\label{eq:energy}
\end{eqnarray}
\end{widetext}
Note that the larger the value of n, the longer the wave train is (the narrower the spectrum is). Now we can explain the experiments shown in Fig.2. We use Eq.(6) to fit the spectra in Fig.2. The results are: (a) $n=1$, $\lambda_0 = 653$ nm, $\Delta \lambda  = 122.3$ nm; (b) $n=20$, $\lambda_0 = 560$ nm, $\Delta \lambda  = 11.4$ nm; (c) $n=5$, $\lambda_0 = 550$ nm, $\Delta \lambda  = 40$ nm; (d) $n=1$, $\lambda_0 = 540$ nm, $\Delta \lambda  = 103.4$ nm; (e) $n=3$, $\lambda_0 = 404.7$ nm, $\Delta \lambda  = 41$ nm. The fitting curves are shown in the red dashed ones in Fig.2. One can see that as long as take the physical quantity $\nu$ in Eq. (2) as the center frequency $\nu_0$ of the wave train Eq. (4), and assumes that each photoelectron can only carry one portion of discrete light energy ${p_{0k}}{\omega _0} = h{\nu _0}$, then the result shown Fig.2 can be understand immediately, since the stopping voltages corresponding to Fig.2(a)-(e) are all determined by the center wavelengths of the spectra of incident lights. Besides, the maximum bandwidth $\Delta \lambda$ of a wave train is determined by the spectrum of Eq. (6) with $n=1$.

Next, let’s look at Fig.3. Although there is only little difference between the experimental spectra of Fig.3(b) and Fig.2(a), we find that it is impossible to fit the experimental spectrum of Fig.3(b) with a single spectrum of Eq.(6). Furthermore, we also find that it is impossible to fit Fig.3 (a) or (c) with a single spectrum of Eq.(6). That is, any of the incident lights in Fig.3 (a)-(c) cannot be taken as a single wave train described by Eq.(4). In fact, they all involve multiple components of wave trains. For example, Fig.3 (b) or (c) involves at least two components of wave trains. The fitting of each spectrum is done with the least components and the results are: (b) $n=1$, $\lambda_0 = 652$ nm, $\Delta \lambda  = 108.2$ nm (orange line) and $n=1$, $\lambda_0 = 658$ nm, $\Delta \lambda  = 132.3$ nm (light blue line); (c) $n=1$, $\lambda_0 = 400$ nm, $\Delta \lambda  = 77$ nm (purple line) and $n=2$, $\lambda_0 = 415$ nm, $\Delta \lambda  = 57.4$ nm (light blue line).The compound fitting curve of each group is shown by the red dotted line in the figure. In these two groups of wave trains, the smaller center wavelengths are 562 nm and 400 nm, respectively, which correspond really to the experimental stopping voltages. The experiment suggests that the stopping voltage is determined by the smallest center wavelength when the incident light involves multiple wave train components with different center wavelengths. Let’s look back at Fig.3(a). We will no longer do spectral decomposition and only need to state that the stopping voltage can be explained by a wave train of $n=1$, $\lambda_0 = 439$ nm, $\Delta \lambda  = 83.9$ nm(see the dotted purple line in Fig.3 (a)); and the spectrum of this wave train is indeed at the purple end of the spectrum of bromine tungsten lamp. Following this law, it is easy to understand the above first step of experiment when multiple quasi-monochromatic lights are incident onto the photoelectric tube simultaneously and the stopping voltage is determined by the light of maximum frequency. 

The experiment shown in Fig.4 supports the law concluded above. Let’s see Fig.4(a). It is obtained by adding the spectra of quasi-monochromatic lights 589.3 nm, 577 nm and 546 nm on the basis of Fig.2(a). And the stopping voltage is completely determined by any of added quasi-monochromatic lights (see Fig.4(c)) [Note that the spectrum in Fig.2(a) also involves three components of 589.3 nm, 577 nm and 546 nm]. This can be understood, since as discussed above, the spectrum Fig.2(a) can be described by a single wave train Esq.(4) of $\lambda_0$
and the $\lambda_0$ corresponds to the stopping voltage. And the spectrum Fig.4 (a) is different. It has more than one wave train component. Each of added wavelengths is shorter than 653 nm, According to above law, the stopping voltage will be determined by the added quasi-monochromatic light. Let's look again at Fig.4 (b). It is obtained by adding the spectra of quasi-monochromatic lights 365 nm, 404.7 nm and 435.8 nm on the basis of Fig.3 (a). And the stopping voltage is completely determined by any of added quasi-monochromatic lights (see Fig.4 (c)). It is different from the situation shown in Fig.3 (a), where the stopping voltage corresponds to a center wavelength of the wave train with  $n=1$, $\lambda_0 = 439$ nm and $\Delta \lambda  = 83.9$ nm[Note that the spectrum of this wave train also involves three components of 365 nm, 404.7 nm and 435.8 nm]. This can also be understood, since each of these three added wavelengths is shorter than 439 nm, According to above law, it will play a decisive role in generating stopping voltage. Figure 4 and its analysis show clearly that the stopping voltage is determined by the smallest center wavelength when the incident light involves multiple wave train components with different center wavelengths. 

\section{Conclusion}
In conclusion, for the photoelectric effect, if the incident light is with only one of wave train component, the stopping voltage is determined by the center wavelength of the wave train; if the incident light involves multiple wave train components, the stopping voltage is determined by the smallest center wavelength of these wave train components. This can be seen especially clearly in both the second and fourth steps of experiments above. That is to say, in photoelectric effect equation ${E_k} = h\nu  - {W_0}$ the physical quantity $\nu$ does not refer to the frequency of a monochromatic light, but represents the center frequency of a wave train spectrum. The bandwidth of spectrum $\Delta \lambda$ corresponding to the center wavelength $\lambda_0$ can vary in a large range and the maximum one is given by the spectrum of Eq.(6) with $n=1$. In the visible and near-infrared regions, the maximum $\Delta \lambda$ can be as large as 122 nm. This should arouse more attention in the study of energy exchange between light and matter.

\section {Acknowledgments}
\begin{acknowledgments}
Thanks to Dr. Wang Fu-Juan for providing the filters, Thanks to Dr. Fang Yi-Zhong for his help in measuring the transmission spectrum of the filters, and Thanks to Dr. Wang Zhen-Cheng for offering convenience in using the experimental setup of the photoelectric effect. 
\end{acknowledgments}

\end{document}